\documentclass{jps-cp}
\usepackage{txfonts} 
\usepackage{bm}
\title{Extracting the $\Lambda$ Polarizing Fragmentation Function from Belle $e^+e^-$ Data Within the TMD Formalism}

\author{Marco \textsc{Zaccheddu}$^{1,2,*}$, Umberto \textsc{D'Alesio}$^{1,2}$, Leonard \textsc{Gamberg}$^{3}$ and Francesco \textsc{Murgia}$^{2}$}

\inst{$^{1}$Dipartimento di Fisica, Università di Cagliari, Cittadella Univ., I-09042 Monserrato (CA), Italy\\
$^{2}$INFN, Sezione di Cagliari, Cittadella Univ., I-09042 Monserrato (CA), Italy \\
$^{3}$Division of Science, Penn State Berks, Reading, PA 19610, USA  \\
$^{*}$Speaker.}

\email{marco.zaccheddu@ca.infn.it}

\recdate{March 7, 2022}

\abst{Data from the Belle Collaboration for associated production (with a light unpolarized hadron) and single-inclusive production of transversely polarized $\Lambda$-hyperons in $e^+e^-$ annihilation processes allowed to extract within a TMD approach, for the first time, the $\Lambda$ polarizing fragmentation function.
Recent theoretical developments on the computation of cross sections for single-inclusive hadron production in $e^+e^-$ annihilation processes, within a Soft Collinear Effective Theory approach, combined with the CSS formalism for the double-hadron production case, can be used to perform a renewed analysis, adopting a full TMD factorization scheme. Preliminary results and a comparison between the two analyses are presented.}

\kword{TMD fragmentation functions, transverse polarization, polarizing FF, CSS evolution equations}

\begin{document}
\maketitle

\section{Introduction}

Recent Belle data \cite{Belle:2018ttu} for the transverse polarization of $\Lambda$'s, produced in $e^+e^-\to \Lambda^{\uparrow} \pi^{\pm}/K^{\pm} +X $ and in $e^+e^-\to \Lambda^{\uparrow}(jet) +X$, allowed to extract for the first time the $\Lambda$ polarizing fragmentation function (pFF) \cite{previous_extr}. Here we present a renewed analysis of Belle data by exploiting the CSS evolution equations and the recent theory developments on the factorization of single-inclusive hadron production in $e^+e^-$ annihilation processes. The main expressions for the polarization of a spin-$1/2$ hadron produced in $e^+e^-\to h_1^{\uparrow} h_2+X $ and in $e^+e^-\to h_1^{\uparrow}(jet)+X$ will be presented, along with the preliminary results for the renewed extraction of the $\Lambda$ pFF.

\section{Formalism}
\subsection{Double-hadron production: Transverse Polarization}
We consider the process  $e^+e^- \to h^{\uparrow}_1 h_2 +X$, where $h_{1,2}$ are hadrons produced almost back-to-back, and, in the \emph{hadron-frame} configuration, $h_1$ has a small transverse momentum, $\bm{P}_{1T}$, with respect to the $h_2$ direction (for full details on the kinematic  configuration see Ref. \cite{2021}).
The transverse polarization for a spin-$1/2$ hadron $h_1$, produced 
with an unpolarized light-hadron $h_2$, as a function of the light-cone momentum fractions, $z_{1,2}$, of the two final state hadrons, is defined as the ratio of two $\bm{q}_T$-integrated convolutions:

\begin{equation}
    P^{h}_n(z_1,z_2) =  \frac{ \int d^2\bm{q}_T \, F^{\sin(\phi_1 - \phi_{S_1})}_{TU} }{ \int d^2\bm{q}_T \, F_{UU}} = \frac{M_{h_1} \int dq_T\;q_T\;d\phi_1 \, \mathcal{B}_1 \Big[\widetilde{D}^{\perp  (1)}_{1T} \widetilde{\bar{D}}_1\Big]}{ \int dq_T\;q_T\;d\phi_1 \, \mathcal{B}_0 \Big[\widetilde{D}_1 \widetilde{\bar{D}}_1\Big] }\,,
\label{pol_ratio}
\end{equation}
where $M_{h_1}$ is the mass of $h_1$, $F^{\sin(\phi_1 - \phi_{S_1})}_{TU}$ and $F_{UU}$ are the convolutions defined in the transverse momentum space \cite{2021}, and  $\bm{q}_T = -\bm{P}_{1T}/z_1$.
The light-cone fraction can be related to the momentum fraction, $z_{p_i}={2|\bm{P}_{h_i}| }/{\sqrt{s}}$, and to the usual energy fraction, $z_{h_i}={2E_{h_i}}/{\sqrt{s}}$, as  $z_{h/p} = z [1 \pm {M^2_{h}}/{( z^2 {Q}^2)}]$.
According to the  TMD factorization theorems and by exploiting the CSS evolution equations, the full form of the $\mathcal{B}_0$ and $\mathcal{B}_1$ convolutions is given by the following expressions:

\begin{equation}
\begin{split}
    \mathcal{B}_0 \Big[\widetilde{D}_1 \widetilde{\bar{D}}_1\Big]
    =& \frac{\mathcal{H}^{(e^+e^-)}(Q)}{z^2_1 z^2_2} \sum_q e^2_q\int \frac{d b_T}{2 \pi} \, b_T J_0(b_T\, q_T) \, d_{q/h_1}(z_1; \bar{\mu}_b)  d_{\bar{q}/h_2}(z_2; \bar{\mu}_b)\, e^{- S_{\text{pert}}\,(\bar{\mu}_{b},Q)}  \\
    \times& M_{D_1}(b_c(b_T);b_{\text{max}}) \, M_{D_2}(b_c(b_T);b_{\text{max}})\exp\Bigg\{-g_K(b_c(b_T);b_{\text{max}})\ln{\bigg(\frac{Q^2 z_1 z_2}{M_{h_1} M_{h_2}}\bigg)}\Bigg\}\,,
\end{split}
\label{B0_full}
\end{equation}

\begin{equation}
\begin{split}
    \mathcal{B}_1 \Big[\widetilde{D}^{\perp  (1)}_{1T} \widetilde{\bar{D}}_1\Big]
    =& \frac{\mathcal{H}^{(e^+e^-)}(Q)}{z^2_1 z^2_2}\sum_q e^2_q\int \frac{d b_T}{2 \pi} \, b^2_T J_1(b_T\, q_T) \widetilde{D}^{\perp  (1)}_{1T} (z_1;\bar{\mu}_b) \, d_{\bar{q}/h_2}(z_2; \bar{\mu}_b)\, e^{- S_{\text{pert}}\,(\bar{\mu}_{b},Q)}   \\
    \times& M^{\perp}_{D}(b_c(b_T);b_{\text{max}}) \, M_{D_2}(b_c(b_T);b_{\text{max}})\exp\Bigg\{-g_K(b_c(b_T);b_{\text{max}})\ln{\bigg(\frac{Q^2 z_1 z_2}{M_{h_1} M_{h_2}}\bigg)}\Bigg\} \,,
\end{split}
\label{B1_full}
\end{equation}
where $\mathcal{H}^{(e^+e^-)}(Q)$ is the hard scattering part, for the massless on-shell process  $e^+e^-\to q \bar{q}$, at the center-of-mass energy $Q$.
By employing the integral definition of the Bessel functions of the first kind $J_{0,1}$, we can see that both convolutions are given  as Fourier transforms in $\bm{b}_T$-space of different elements:  $d_{q/h_i}$ are the unpolarized collinear fragmentation functions (FFs), $\widetilde{D}^{\perp  (1)}_{1T}$ is the first moment of the polarizing fragmentation function, $M_{D_i}$ and $ M^{\perp}_{D}$  are, respectively, the parameterizations of the non-perturbative functions for the unpolarized hadron and for the transversely polarized hadron, and they have to be extracted from data. $g_K$ is a universal non-perturbative function, that takes into account the effect of the recoil against the emission of soft gluons and, lastly, $S_{\text{pert}}$ is the perturbative Sudakov form factor defined as follows:

\begin{equation}
    S_{\text{pert}}\,(\bar{\mu}_{b},Q) = -\widetilde{K}(b_*;\bar{\mu}_{b}) \ln\frac{Q^2}{\bar{\mu}_{b}^2} - \int^{Q}_{\bar{\mu}_{b}} \frac{d\mu'}{\mu'}\,\bigg[ 2\gamma_D(g(\mu');1) - \gamma_K(g(\mu')) \ln{\frac{Q^2}{\mu'^2}} \bigg]\,.
\label{S_pert}
\end{equation}

\noindent The integration over the azimuthal dependence $d\phi_1 $ in Eq. (\ref{pol_ratio}) gives a factor of $2\pi$ that cancels in the polarization definition. %
Meanwhile for the radial part, since the only terms inside the convolutions that depend on $q_T $ are the Bessel functions, we can separately integrate them, obtaining the following results:
\begin{equation}
    \begin{split}
    \int^{q_{T_{\text{max}}}}_0 dq_T \, q_T J_0(b_T\, q_T) =& \frac{q_{T_{\text{max}}}}{b_T}J_1(b_T\, q_{T_{\text{max}}}) \\
    \end{split}
\label{int_qtmax_1}
\end{equation}
\begin{equation}
    \begin{split}
    &\int^{q_{T_{\text{max}}}}_0 dq_T \, q_T J_1(b_T\, q_T) =\frac{\pi q_{T_{\text{max}}}}{2 b_T}
    \bigg\{J_1(b_T\, q_{T_{\text{max}}})\bm{H}_0(b_T\, q_{T_{\text{max}}}) - J_0(b_T\, q_{T_{\text{max}}})\bm{H}_1(b_T\, q_{T_{\text{max}}}) \bigg\} \,,
    \end{split}
\label{int_qtmax_2}
\end{equation}
where ${\bm{H}}_{0,1}$ are the Struve functions of order zero and one respectively. To fulfill the conditions of validity for the TMD factorization we require that $q_{T_{\text{max}}} \ll Q$.


\subsection{Single-inclusive hadron production: Transverse polarization}
In a similar manner, it is possible to define the transverse polarization for a spin-$1/2$ hadron, produced in $e^+e^- \to h^{\uparrow}_1(jet) +X$, as:
\begin{equation}
    \mathcal{P}(z_1,j_{\perp}) = \frac{{d\Delta \sigma}/{dz_1 d^2\bm{j}_{\perp}}}{{d \sigma}/{dz_1 d^2\bm{j}_{\perp}}}\,,
\label{eq:1_h_polarization}
\end{equation}
where $\bm{j}_{\perp}$ is the hadron transverse momentum measured with respect to the thrust axis. The polarization is defined again as the ratio of two convolutions, derived in \cite{kang}, whose full forms are given by the following Fourier transforms (with $\bm{q}_T= - \bm{j}_{\perp}/z_1 $):

\begin{equation}
\begin{split}
    \frac{d \sigma}{dz_1 d^2 \bm{j}_{\perp}} =& \frac{\sigma_0}{z^2_1} \sum_q e^2_q\int \frac{d b_T}{2 \pi} \, b_T J_0(b_T\, q_T)d_{q/h_1}(z_1; \bar{\mu}_b) \, U_{NG}(\bar{\mu}_b,Q) \\
    &\times M_{D_1}(b_c(b_T);b_{\text{max}}) \, \exp\Bigg\{-g_K(b_c(b_T);b_{\text{max}})\ln{\bigg(\frac{Q z_1}{M_{h_1}}\bigg)}\Bigg\}\, e^{- S_{\text{pert}}\,(\bar{\mu}_{b},Q)/2},
\end{split}
\label{eq:1-h_unp}
\end{equation}

\begin{equation}
\begin{split}
    \frac{d \Delta\sigma}{dz_1 d^2\bm{j}_{\perp}} =&\frac{\sigma_0}{z^2_1} \sum_q e^2_q\int \frac{d b_T}{2 \pi} \, b^2_T J_1(b_T\, q_T) \widetilde{D}^{\perp (1)}_{1T}(z_1,\bar{\mu}_b) \, U_{NG}(\bar{\mu}_b,Q) \\
    &\times M^{\perp}_{D}(b_c(b_T);b_{\text{max}})  \, \exp\Bigg\{-g_K(b_c(b_T);b_{\text{max}})\ln{\bigg(\frac{Q z_1}{M_{h_1}}\bigg)}\Bigg\}\, e^{- S_{\text{pert}}\,(\bar{\mu}_{b},Q)/2}.
\end{split}
\label{eq:1-h_pol}
\end{equation}
The same elements, entering the double-hadron production case, also appear here: namely $M^{\perp}_{D}$, up to NLO calculations, as well as the first moment of the pFF, $\widetilde{D}^{\perp (1)}_{1T}$.
Moreover, since soft radiation is restricted to only one hemisphere, the cross section is a non-global observable,  which was also used in the study \cite{Gamberg:2021iat}. In both convolutions, Eqs.~(\ref{eq:1-h_unp}) and (\ref{eq:1-h_pol}), the function $ U_{NG}$, accounts for the effect of such non-global logarithms. We employ the parametrization given in Ref.~\cite{Dasgupta:2001sh}:

\begin{equation}
    U_{NG}(\mu_{b_*},Q) = \exp{\bigg[ -C_A C_F \frac{\pi^2}{3} u^2 \frac{1+ (au)^2}{1+(bu^c)} \bigg]}\,,
\end{equation}
with $a = 0.85 \, C_A$, $b = 0.86 \, C_A$, $c = 1.33$ and
\begin{equation}
    u = \frac{1}{\beta_0} \ln{\bigg[\frac{\alpha_s(\mu_{b_*})}{\alpha_s(Q)} \bigg]}\,,
\end{equation}
where $\beta_0 = \frac{11}{3}C_A - \frac{4}{3} T_F n_f$, with $T_F = 1/2$ and $n_f$ is the number of the active flavors.

\section{Phenomenology}
\begin{figure}[h!]
\includegraphics[width=8cm]{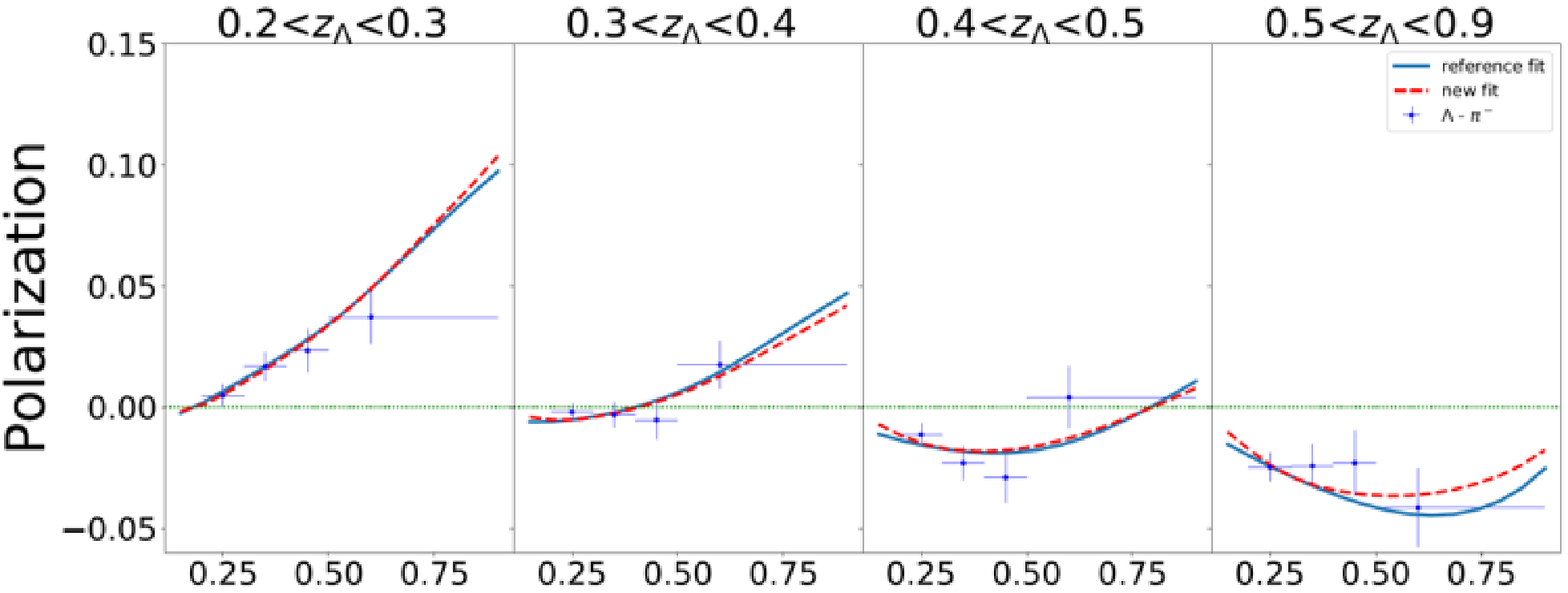}
\includegraphics[width=8cm]{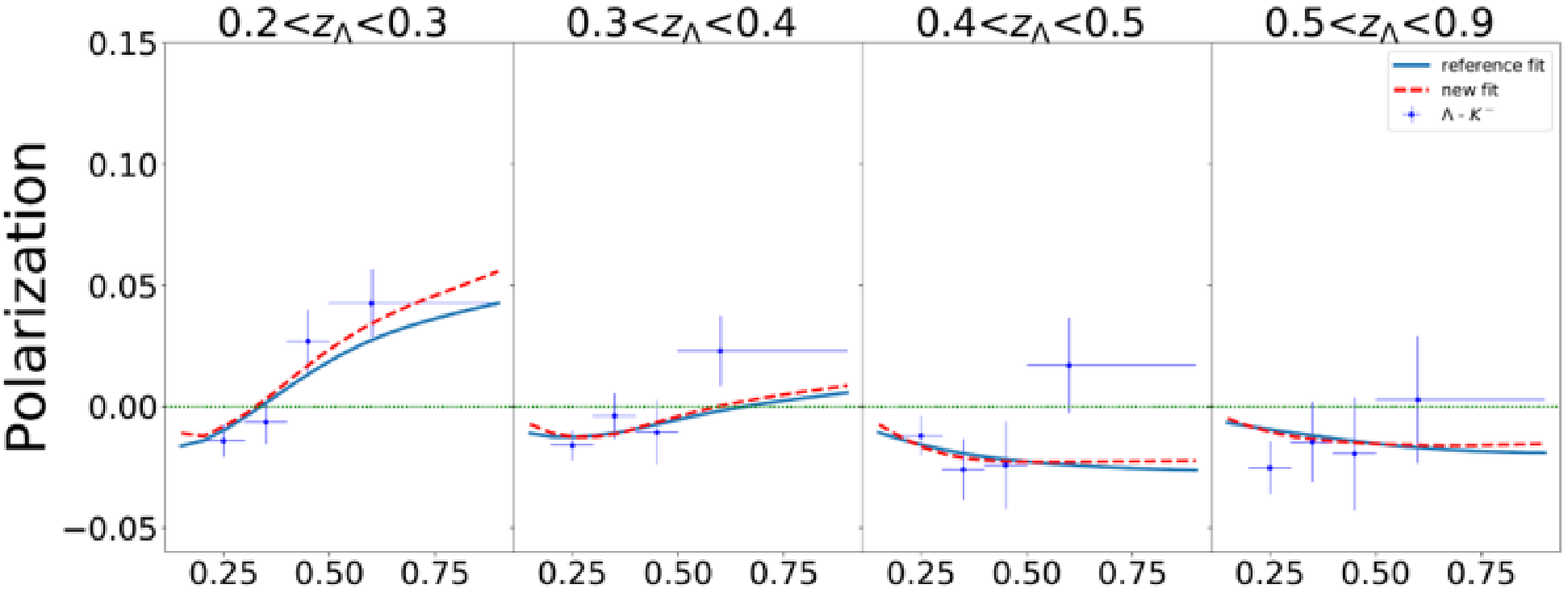}
\includegraphics[width=8cm]{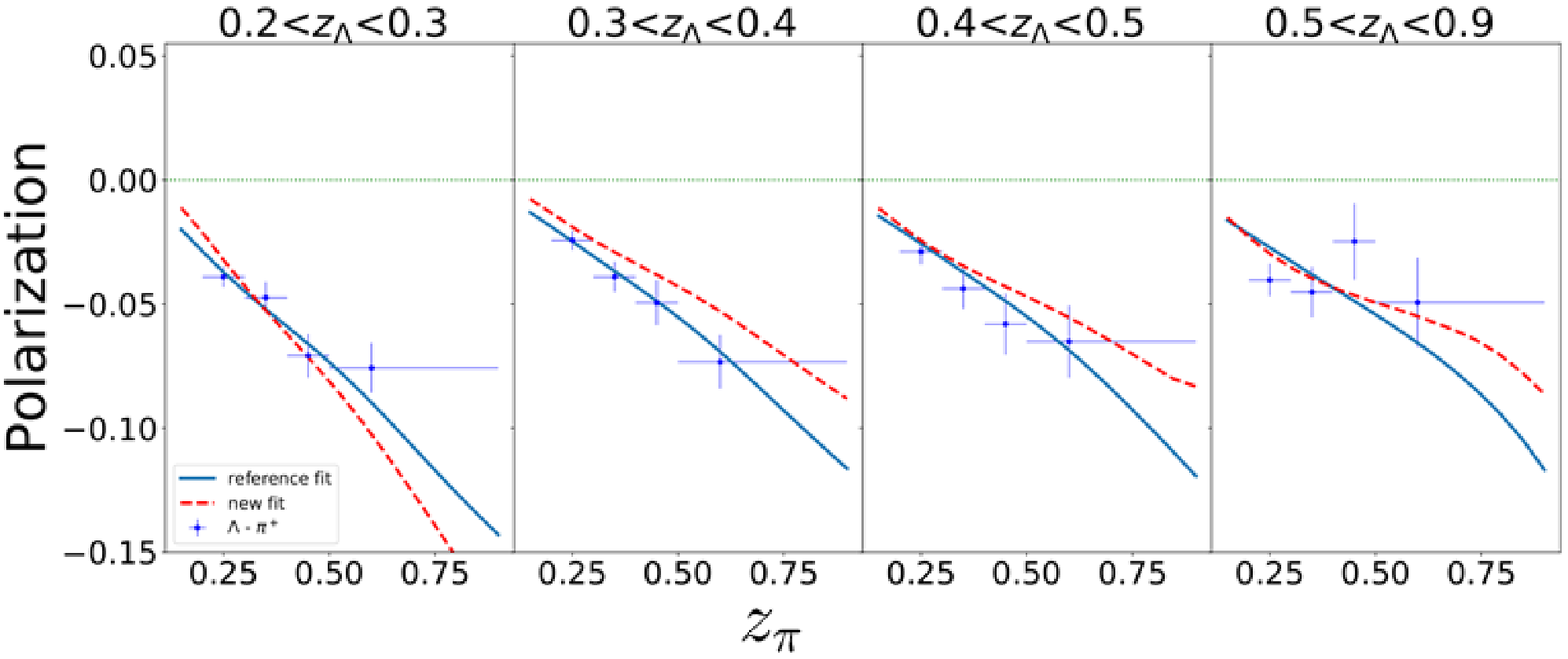}
\includegraphics[width=8cm]{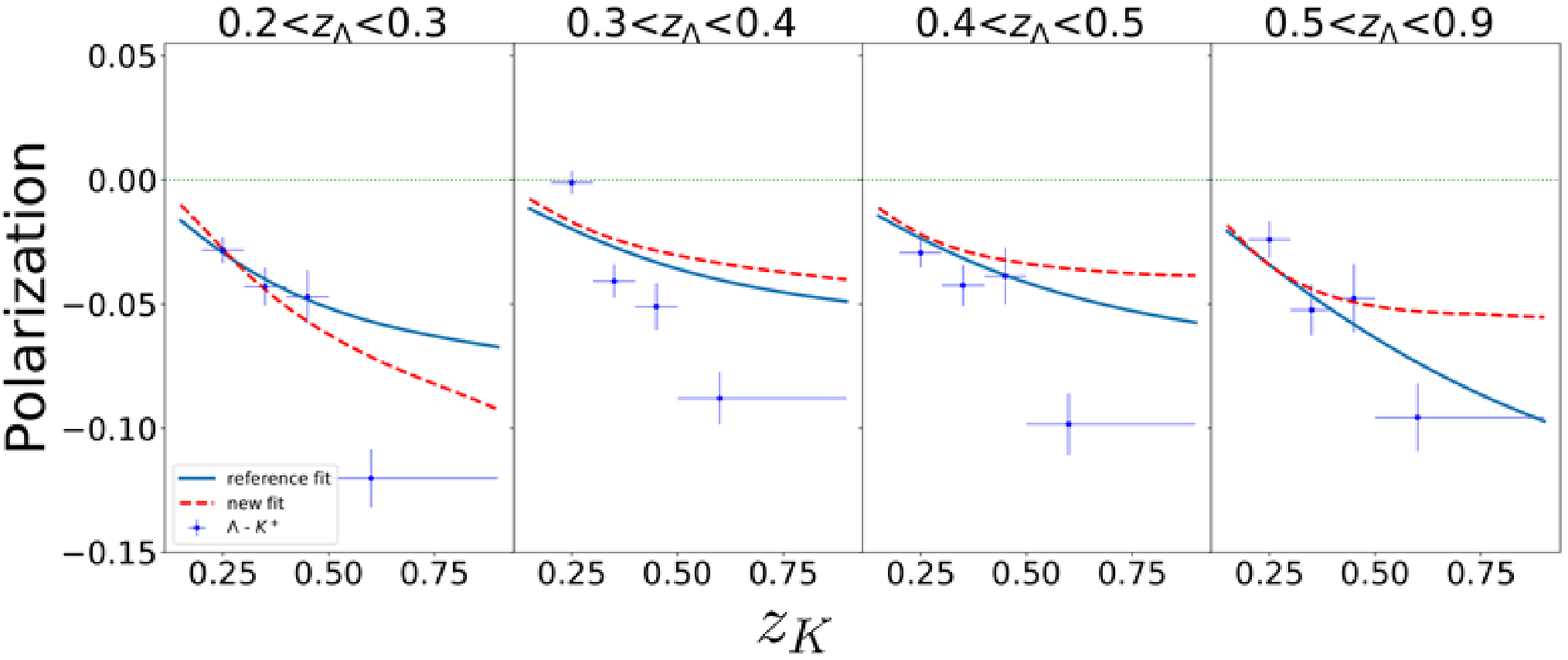}
\caption{Best-fit estimates (red dashed lines) of the transverse polarization for $\Lambda$'s produced in $e^+e^-\to \Lambda h +X $, for $h =\pi^{\pm}, K^{\pm}$, as a function of $z_h$ (of the associated hadron) for different $z_{\Lambda}$ bins, obtained using the Power-Law model and compared to our previous best-fit estimates (blue solid lines)\cite{previous_extr}. Data are from Belle \cite{Belle:2018ttu}.}
\label{f1}
\end{figure}

We first perform a fit of the double-hadron production data alone, in order to extract $\widetilde{D}^{\perp  (1)}_{1T,\, \Lambda/q}$ and $M^{\bot}_{D, \Lambda}$.
The 126 Belle data points are given as a function of $z_{\Lambda}$ and $z_{\pi/K}$, the energy fractions of the $\Lambda$/$\bar{\Lambda}$ and $\pi/K$ particles. By imposing  a cut for large values of the light-hadron energy fractions, $z_{\pi/K}<0.5$, we are left with only 96 data points.
The first moment is parametrized as follows ($q=u,d,s,\text{sea}$):

\begin{equation}
    \begin{split}
        \widetilde{D}^{\perp  (1)}_{1T,\, \Lambda/q}(z;\mu_b)=\mathcal{N}^p_q(z)d_{q/\Lambda}(z;\mu_b) \\
        \mathcal{N}^p_q(z) = N_q z^{a_q}(1-z)^{b_q}\frac{(a_q +b_q )^{(a_q +b_q )}}{a_q^{a_q}b_q^{b_q}}\,.
    \end{split}
\end{equation}
The best parameter choice is given by: $N_u,\, N_d,\, N_s,\, N_{\text{sea}},\, a_s,\, b_u,\, b_{\text{sea}} $, with all other $a$ and $b$ parameters set to zero. Regarding the non-perturbative functions $M_D$ and $g_K$, we have used various combinations of them, and the best results are found by employing the following expressions (see also \cite{coll_rog}):

\begin{equation}
    \begin{split}
    %
    M_D(b_c(b_T);b_{\text{max}}) &= \exp{\bigg(-\frac{\langle p_\perp^2 \rangle b^2_T}{4 z^2_p}\bigg)} \qquad \text{Gaussian} \\
    %
    %
    M_D(b_c(b_T);b_{\text{max}}) &= \frac{2^{2-p}}{\Gamma(p-1)}\,(b_T m/z_p)^{p-1}{K}_{p-1}(b_T m/z_p)\qquad \text{Power-Law (2)} \,,\\
    \end{split}
\end{equation}
\begin{equation}
    \begin{split}
        %
        %
        g_K(b_c(b_T);b_{\text{max}}) &= \frac{\alpha_s(C_1/b_*) C_F}{\pi} \ln(1+ b^2_T/b^2_{\text{max}})  \qquad \text{Logarithmic}\\
        %
        %
    \end{split}
\end{equation}
\begin{table}[tbh]
\centering
\caption{Best values of the $\chi^2_{\text{dof}}$ obtained fitting the double-hadron production data set only.}
\label{t1}
\begin{tabular}{c c c c c}
\hline \hline
Polarizing & Unpolarized & $g_K$ & $q_{T_{\text{max}}}/Q$ & $\chi^2_{\text{dof}}$ \\[.7ex]
\hline
Power-Law (2)  & Power-Law (2) & Logarithmic & $0.25$& $1.203$ \\
Power-Law (2) & Gaussian & Logarithmic & $0.25$ & $1.211$ \\
\hline \hline
\end{tabular}
\end{table}
leading to $\chi^2_{\text{dof}} \simeq 1.2$. In Tab. \ref{t1} we report the exact $\chi^2_{\text{dof}}$ values according to the combination of the non-perturbative functions employed for the unpolarized and polarizing FFs, and to the $q_{T_{\text{max}}}/Q$ value used.  Depending on the non-perturbative functions utilized, 8 or 9 parameters have been extracted. In addition, reasonable $\chi^2_{\text{dof}}$ values ($\chi^2_{\text{dof}}<1.37$) can also be obtained  with other combinations of the non-perturbative functions, allowing for a good description of the experimental data.
In Fig.~\ref{f1} we show the estimates, obtained adopting the parameters extracted with the Power-Law model (2) (see first row of Tab. \ref{t1}), of the transverse polarization of $\Lambda$'s , produced in association with a light-hadron, compared to Belle data~\cite{Belle:2018ttu} and to our previous fit results \cite{previous_extr}.
We then perform a combined fit, including also the single-inclusive $\Lambda$ production data, on which we impose a cut for small values of the $\Lambda$ energy fraction, $z_{\Lambda}<0.3$. This leads to a total of 119 data points. A good description is obtained by employing the following non-perturbative functions:

\begin{equation}
    \begin{split}
    %
    %
    M_D(b_c(b_T);b_{\text{max}}) &= \frac{2^{2-p}}{\Gamma(p-1)}\,(b_T m)^{p-1}{K}_{p-1}(b_T m)\qquad \text{Power-Law (1)}\\
    %
    \end{split}
\end{equation}
\begin{equation}
    \begin{split}
        g_K(b_c(b_T);b_{\text{max}}) = g_2 \ln\bigg(\frac{b_T}{b_*}\bigg) \quad g_2 = 0.84  \qquad \text{AFGR / SIYY}\quad \text{\cite{afgr,siyy}}\,.
    \end{split}
\end{equation}

\begin{figure}[h!]
\centering
\includegraphics[width=10cm]{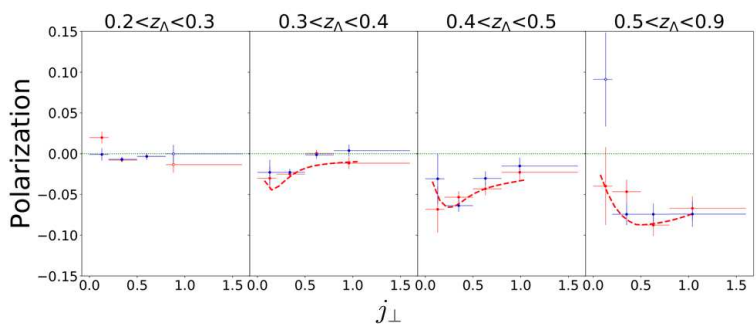}
\caption{Best-fit estimates of the transverse polarization for $\Lambda$'s produced in  $e^+e^-\to \Lambda(jet) +X $, obtained using the Power-Law model, as a function of the transverse momentum $j_{\perp}$ and for different $z_{\Lambda}$ bins.}
\label{f2}
\end{figure}
As reported in Tab. \ref{t2} the combined fit leads to a $\chi^2_{\text{dof}} = 1.58$. In Fig. \ref{f2} we show the estimates for the transverse polarization against Belle data. As we can see, we are able to describe the size and the pattern of the data, showing the possibility to perform a combined fit of both data sets, and obtaining a better description as compared to our previous analysis of the single-inclusive data set \cite{previous_extr}.

\begin{table}[tbh!]
\centering
\caption{Value of the $\chi^2_{\text{dof}}$ obtained with the combined fit of both the double-hadron and the single-inclusive hadron production data sets.}
\label{t2}
\begin{tabular}{c c c c c}
\hline \hline
Polarizing & Unpolarized & $g_K$ & $q_{T_{\text{max}}}/Q$ & $\chi^2_{\text{dof}}$ \\[.7ex]
\hline
Power-Law (1) & Gaussian & AFGR/SIYY & $0.25$& $1.58$ \\
\hline \hline
\end{tabular}
\end{table}

\section{Conclusions}

The main expressions for the transverse polarization of a spin-$1/2$ hadron have been presented, both for double and single-inclusive hadron production in $e^+e^-$ annihilation processes, in terms of convolutions in $\bm{b}_T$-space, by employing the CSS evolution equations. We have then illustrated the analyses of two fits, adopting different combinations of the non-perturbative functions: one restricted to the double-hadron production case and one including also the single-inclusive hadron production. A comparison with a previous extraction has been discussed.

\end{document}